\documentclass[9pt,sigconf,letterpaper]{acmart}
\usepackage[utf8]{inputenc}
\usepackage{graphicx}
\usepackage{subfigure}
\usepackage{array}
\usepackage{booktabs}
\usepackage[disable]{todonotes} %
\usepackage{balance}
\usepackage{colortbl}
\definecolor{ColorAtkNo}{rgb}{0.59,0.73,0.38}
\definecolor{ColorAtkYes}{rgb}{0.97,0.22,0.10}

\usepackage[absolute,showboxes]{textpos}

\usepackage{sistyle}

\acmYear{2021}\copyrightyear{2021}
\setcopyright{acmcopyright}
\acmConference[IMC '21]{ACM Internet Measurement Conference}{November 2--4, 2021}{Virtual Event, USA}
\acmBooktitle{ACM Internet Measurement Conference (IMC '21), November 2--4, 2021, Virtual Event, USA}
\acmPrice{15.00}
\acmDOI{10.1145/3487552.3487840}
\acmISBN{978-1-4503-9129-0/21/11}


\newcommand{\result}[1]{}

\definecolor{myred}{cmyk}{0, 0.7808, 0.4429, 0.1412}

\newcommand{\done}[1]{}

\usepackage{pifont}
\newcommand{\cmark}{\ding{51}}%
\newcommand{\xmark}{\ding{56}}%

\usepackage{xspace}
\newcommand{\etal}{\textit{et al.}~}
\newcommand{\eg}{\textit{e.g.,}~}
\newcommand{\ie}{\textit{i.e.,}~}

\newcommand{\one}{({\em i})\xspace}
\newcommand{\two}{({\em ii})\xspace}
\newcommand{\three}{({\em iii})\xspace}

\makeatletter
\renewcommand{\paragraph}[1]{\vspace*{0.03in}\noindent{\bf #1.}\hspace{0.25ex \@plus1ex \@minus.2ex}}
\newcommand{\paragraphNoDot}[1]{\vspace*{0.03in}\noindent{\bf #1}\hspace{0.25ex \@plus1ex \@minus.2ex}}
\makeatother

\begin{document}

\date{}

\setlength{\TPHorizModule}{\paperwidth}
\setlength{\TPVertModule}{\paperheight}
\TPMargin{5pt}
\begin{textblock}{0.8}(0.1,0.02)
	\noindent
	\footnotesize
	\centering
	If you cite this paper, please use the IMC reference:
	M. Nawrocki, R. Hiesgen, T. C. Schmidt, and M. Wählisch.
	2021. QUICsand: Quantifying QUIC Reconnaissance Scans and DoS Flooding Events.
	\emph{In Proceedings of ACM Internet Measurement Conference (IMC ’21).}
	ACM, New York, NY, USA, 7 pages. https://doi.org/10.1145/3487552.3487840
\end{textblock}

\title[QUICsand: Quantifying QUIC Reconnaissance Scans and DoS Flooding Events]{QUICsand: Quantifying QUIC Reconnaissance Scans and \\ DoS Flooding Events}

\begin{abstract}

In this paper, we present first measurements of Internet background radiation originating from the emerging transport protocol QUIC.
Our analysis is based on the UCSD network telescope, correlated with active measurements.
We find that research projects dominate the QUIC scanning ecosystem but also discover traffic from non-benign sources.
We argue that although QUIC has been carefully designed to restrict reflective amplification attacks, the QUIC handshake is prone to resource exhaustion attacks, similar to TCP SYN floods.
We confirm this conjecture by showing how this attack vector is already exploited in multi-vector attacks:
On average, the Internet is exposed to four QUIC floods per hour and half of these attacks occur concurrently with other common attack types such as TCP/ICMP floods.

\end{abstract}

\author{Marcin Nawrocki}
\email{marcin.nawrocki@fu-berlin.de}
\affiliation{%
  \institution{Freie Universit{\"a}t Berlin}
  \country{Germany}
}

\author{Raphael Hiesgen}
\email{Raphael.Hiesgen@haw-hamburg.de}
\affiliation{%
  \institution{HAW Hamburg}
  \country{Germany}
}

\author{Thomas C. Schmidt}
\email{t.schmidt@haw-hamburg.de}
\affiliation{%
  \institution{HAW Hamburg}
  \country{Germany}
}

\author{Matthias W\"ahlisch}
\email{m.waehlisch@fu-berlin.de}
\affiliation{%
  \institution{Freie Universit{\"a}t Berlin}
  \country{Germany}
}

\renewcommand{\shortauthors}{Nawrocki et al.}

\maketitle

\section{Introduction}
\label{sec:introduction}

QUIC is a secure transport protocol originally developed by Google and tested in Chrome browsers since 2013 \cite{Langley2017Quic}.
It has been recently standardized by the IETF as RFC 9000 \cite{RFC-9000} and at the same time enjoys rapidly growing deployment by major Web operators and browsers.
In 2017, Google estimated that QUIC accounted for 7\% of Internet traffic \cite{Langley2017Quic} and, by the end of 2020, Facebook announced that 75\% of its Internet traffic is QUIC \cite{Facebook2020Quic}.
Despite its recent standardization, QUIC has already many implementations~\cite{Piraux2018Evolution} and concurrently supported QUIC versions~\cite{rueth2018quic}.
In 2021, scans of the complete IPv4 address space  detected around 2 million QUIC servers~\cite{rueth2018quic}.

Key design objectives in QUIC were privacy and security. QUIC was built to reduce the attack surface on the transport layer, which includes attacks such as reflective amplifications~\cite{rowrs-adads-15} and resource exhaustions. Security considerations in the QUIC RFC ~\cite{RFC-9000} span 18~pages and discuss properties against active and passive~attackers.

In this paper, we report about early observations that indicate regularly ongoing attacks based on QUIC.
We argue that the strong security model in QUIC does not preclude misuse and measure clear signals of DDoS attacks in Internet background radiation. We confirm these results by correlating our observations with several complementary
data sources.
Our findings indicate that QUIC servers are indeed prone to resource exhaustion attacks and these flaws are currently exploited in multi-vector attacks.   
We believe that it will be crucial to monitor such attack attempts  early in the QUIC deployment phase before they enfold their full potential.

The main contributions of our paper summarize as follows:
\begin{enumerate}
\item We present the first study on QUIC Internet background radiation as seen by a large network telescope.
\item We show a significant bias by research scanners but also detect scanning activity from non-benign sources.
\item Surprisingly, we find high-volume backscatter events suggesting that QUIC is used in multi-vector resource exhaustion attacks, targeting well-known companies.
\item We benchmark a popular web server implementation to test its DoS resiliency and validate our observations.
\item We show the efficacy of QUIC's built in defense mechanism with \texttt{RETRY} messages, which remains unused in the wild based on our measurements.
\end{enumerate}

In the remainder of this paper, we present background and related work in \autoref{sec:related_work}.
We outline the QUIC attack scenarios that we base on in \autoref{sec:attackmodel}, and introduce our measuement method and data sources in \autoref{sec:method}.
We analyse QUIC scanning and backscatter events in \autoref{sec:analysis}.
Finally, we discuss our findings in \autoref{sec:discussion}. %

\section{Background and Related Work}
\label{sec:related_work}

One of the major design goals in QUIC is the decrease of latencies for client-server applications by reducing round-trip times (RTTs) due to multiple, independent handshakes~\cite{Cui2017Quic}.
To this end, sequential handshakes from TCP, TLS, and HTTP have been merged into a single, comprehensive handshake process.
Furthermore, to overcome delay because of TCP head-of-line blocking, QUIC is based on UDP \cite{Langley2017Quic}.

\begin{figure}[t]
  \begin{center}
  \includegraphics[width=1\columnwidth]{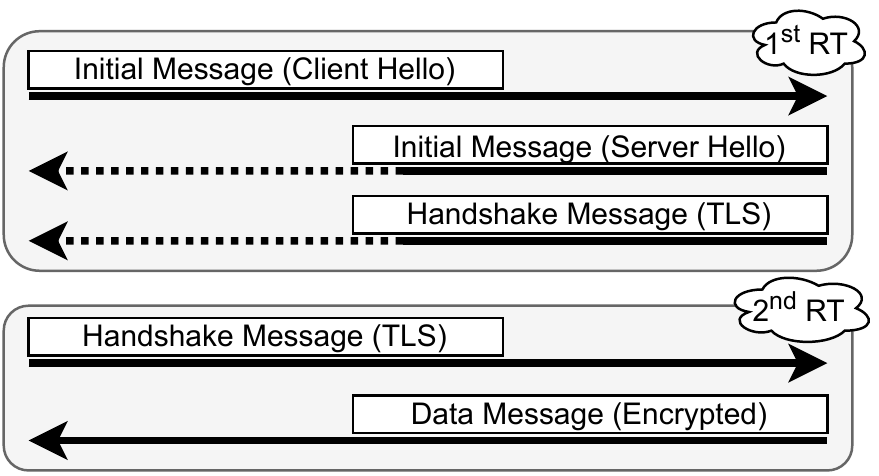}
      \caption{QUIC clients receive data with the second round-trip (RT). At the first RT, servers respond to unverified client IP addresses, which can be misused.}
  \label{fig:quic_handshake}
  \end{center}
\end{figure}

\paragraph{QUIC handshakes}
QUIC utilizes various handshake procedures to set up and resume connections.
In the best case, when cached information of a prior connection is available to the client, encrypted application data can be sent immediately leading to a 0-RTT handshake.
In the worst case, the handshake requires 3 RTTs \cite{rueth2018quic}:
If the client offers unsupported QUIC versions, the server first enforces a version negotiation and then proceeds with a \textit{typical} \cite{Gagliardi2020Sessions} handshake.

A typical handshake is depicted in \autoref{fig:quic_handshake}.
First, the client sends an \texttt{Initial} message including a \texttt{TLS Client Hello}, which is answered with an \texttt{Initial} message including a \texttt{TLS Server Hello}.
This message is immediately followed by a \texttt{handshake} message containing the rest of the TLS server information (\eg certificates).
The setup is complete with a \texttt{Handshake} message from the client and is usually accompanied by a data request.
Since the client is allowed to send data requests after the first round-trip, this handshake is described as a 1-RTT handshake.
During the handshake, the server and client agree on a \one source connection identifier (SCID) and a \two destination connection identifier (DCID), respectively, which can be used to identify the QUIC connection independently of the traditionally used connection 5-tuple.

\paragraph{\texttt{RETRY} to mitigate resource exhaustion}
QUIC traffic is almost entirely encrypted with TLS 1.3 to prevent the ossification that middleboxes cause on protocols like TCP \cite{Honda2011Extend}.
During the first RTT, a server responds to an unverified client IP address.
This means that the server performs cryptographic operations for a potentially spoofed client.
QUIC supports \texttt{RETRY} messages \cite{RFC-9000} to limit the attack surface of resource exhaustion attacks.
\texttt{RETRYs} precede a typical QUIC handshake and force the client to respond with a unique token, which proves its authenticity.
This mitigation, however, adds a complete RTT, which conflicts with QUIC original design goals.
Recent QUIC server implementations such as NGINX or Picoquic support \texttt{RETRY} \cite{Huitema2020DDoS} but based on the backscatter we observe  \texttt{RETRY} seems rarely deployed.

\paragraph{Related work}
Prior work focusses on three aspects.
First, the performance benefits of QUIC, especiially in low bandwidth, high latency, and high loss~\cite{Biswal2016Faster, Cook2017Better, Carlucci2015HTTP, Kakhki2017Long} or multi-hop~\cite{Coninck2017Multipath} scenarios.
Second, the adoption of QUIC~\cite{rueth2018quic,mtmbb-aaqmd-20,rgdmm-fyewi-20,smp-wfaq-21}.
Third, protocol security.
Previous results~\cite{Lychev2015Secure} suggest that QUIC's security weaknesses, such as insufficient forward secrecy or susceptibility to replay attacks, are introduced by the mechanisms used to reduce latency.
To the best of our knowledge, this is the first paper that analyzes QUIC background radiation and reveals QUIC DoS traffic, showing that we still face a trade-off between small latencies and robust security guarantees in the wild.

\section{QUIC Attack Scenarios}
\label{sec:attackmodel}

In this section, we briefly discuss those attacks that are most relevant in the context of this paper.
An attacker has two common options~\cite{mirkovic2004taxonomy} to exploit weaknesses in QUIC~\cite{Nexusguard2020Amplification, cloudflareBlogQuic}.
First, an attacker could initiate state-overflow attacks to harm QUIC servers.
Second, an attacker could trigger reflective amplification attacks to harm the network.

\paragraph{State-overflow attacks}
To launch a state-overflow attack, an attacker would act as a QUIC client and induce connection states at a QUIC server, by initiating full hand\-shakes.
Any QUIC server answers to the connecting client with a unique Source Connection ID (SCIDs) and its TLS certificate.
This part introduces cryptographic load on the server and forces the allocation of resources to maintain states.
It is noteworthy that during the first round-trip (RT) of the full handshake the client is still unverified, which limits protection mechanisms in QUIC.
To artificially increase the number of states, an attacker would randomly spoof the source IP addresses, source ports, or SCIDs.
Floods benefit from spoofed IP addresses in particular as the backtracking of spoofed traffic is challenging~\cite{lichtblau2017spoofing,ehsw-rssdi-19,mlhcb-cisti-19,ocfmj-tdssi-20}.

This attack is very similar to TCP~SYN~floods~\cite{RFC-4987}, which exploit the fact that the network stack needs to maintain all currently active connections.
Spoofed connections fill up the connection queues at the victim and cause the rejection of legitimate TCP requests.

A large content network has started to mitigate QUIC floods \cite{cloudflareBlogQuic} but the extent of this attack in the wild has not been studied yet.

\paragraph{Reflective amplification attacks}
Since QUIC is based on UDP an attacker could easily launch a reflective amplification attack.
An attacker would send an \texttt{Initial} packet including a spoofed source IP~address to a QUIC server.
The server then replies with an \texttt{Initial} QUIC message and a TLS handshake.
The TLS handshake is larger compared to the client's request, since it includes the server certificates.
As long as the client is unverified (\ie the client did not yet sent messages that include information supplied by the server, see the QUIC RFC~\cite{RFC-9000}) QUIC servers are only allowed to triple the bytes of a request in their response, though.

To increase the absolute number of bytes sent to the (spoofed) victim, an attacker would add padding bytes to the \texttt{Initial} message.
Such strategy would not be suspicious.
Sending large initial handshake packets is suggested in order to allow the server to accommodate full certificates in a single message and thus reducing delays~\cite{Fastly2020Compression}.

\paragraph{Why amplification attacks are unlikely}
QUIC has been designed based on experiences with amplification attacks.
Thus, it limits the size of replies to unverified clients to a factor of \texttt{3$\times$}.
The major reason why QUIC is unlikely to be used for amplification attacks is the wide presence of other protocols that support much higher amplification factors (\eg NTP~\texttt{500$\times$} and DNS \texttt{60$\times$}~\cite{r-ahrnp-14}).
Given TLS certificate compression~\cite{RFC-8879}, only 1\%-9\% of the server replies do not fit in a single message, depending on the initial client message size~\cite{Fastly2020Compression}.
Furthermore, attackers tend to reuse their existing attack infrastructure and adapt very slowly to new protocols to conduct reflective amplification attacks \cite{kopp2021ddos}.
As QUIC amplification attacks are currently unlikely, we focus on state-overflow attacks. %

\section{Measurement Method and Setup}
\label{sec:method}

We analyze QUIC background radiation, and in particular backscatter traffic, \ie responses to spoofed packets.

\subsection{Method}

Our vantage point is a network telescope, since network telescopes passively collect unsolicited traffic known as Internet background radiation (IBR). 
IBR consists of traffic resulting from research and malicious scans~\cite{Richter2019Scanning, msc-ccssv-02}, misconfigurations and bugs but also from responses sent by the victims of randomly spoofed, state-building attacks.
Network telescopes have been used reliably to quantify DoS victims for more than 15 years \cite{moore2006inferring,jonker2017millions}.

We identify QUIC traffic based on transport layer properties by selecting all UDP packets with a source or destination port \texttt{UDP/443}.
This port-based classification has been proven sufficient in prior work \cite{rueth2018quic}.
To exclude false positives, we extend this common classification method by utilizing Wireshark payload dissectors, which also have been shown to be efficient for classification~\cite{nsw-icpic-21}.
We  verified  QUIC detection by manually checking QUIC connections to well-known QUIC servers.
All packets were correctly identified and dissected by Wireshark. 

To group packets sent to the telescope into backscatter and scanning events, we mark all QUIC packets with source port \texttt{UDP/443} as responses (\ie backscatter) and all packets with destination port \texttt{UDP/443} as requests (\ie scans).
These two sets are disjoint, as we do not find any packet with destination and source port set to~\texttt{UDP/443}.

To find DoS events, we apply the notion of traffic sessions and DoS thresholds as defined by Moore \etal \citep{moore2006inferring}.
We acknowledge that these thresholds have been defined with the help of older traffic patterns, however, they are still used in recent work \cite{jonker2017millions}.
Along the line of our analysis we will learn that these thresholds (\eg the \textit{timeout} parameter) are still appropriate for current traffic patterns.
For details on the impact of the threshold configurations, see \autoref{sec:dropped-backscatter}.

\subsection{Data Sources}

We utilize the UCSD Network Telescope \cite{caidaWebsiteTelescope} to observe both QUIC~IBR and common TCP/ICMP IBR.
This telescope is operated by the University of California San Diego and represents a \texttt{/9} network prefix., \ie $1/512$ of the available IPv4 address space.
We are thus able to capture at least 2 \textperthousand\ of any horizontal scan or randomly spoofed attack.
In this paper, we focus on one month, April~1-30,~2021.
Overall, we find 92 million QUIC packets during our measurement period.

In order to bolster our results and put them into context, we correlate our observations with several complementary data sources.
We reuse data from active scans that explore QUIC servers~\cite{rueth2018quic}, correlate IP addresses with the GreyNoise Honeypot Platform, and use meta data from PeeringDB.
The active scans provide a set of potential victims, GreyNoise helps to assess multi-vector attacks based on advanced threat intelligence, and PeeringDB provides additional network descriptions.
We mention that all data sets are in sync, spanning the same period of time.

\section{Analysis}
\label{sec:analysis}

We now analyze QUIC IBR traffic.
First, we show an overview of all IBR traffic and then focus on high-volume backscatter events.

\subsection{Overview of QUIC IBR Traffic}

\begin{figure}[t]
  \begin{center}
  \includegraphics[width=1\columnwidth]{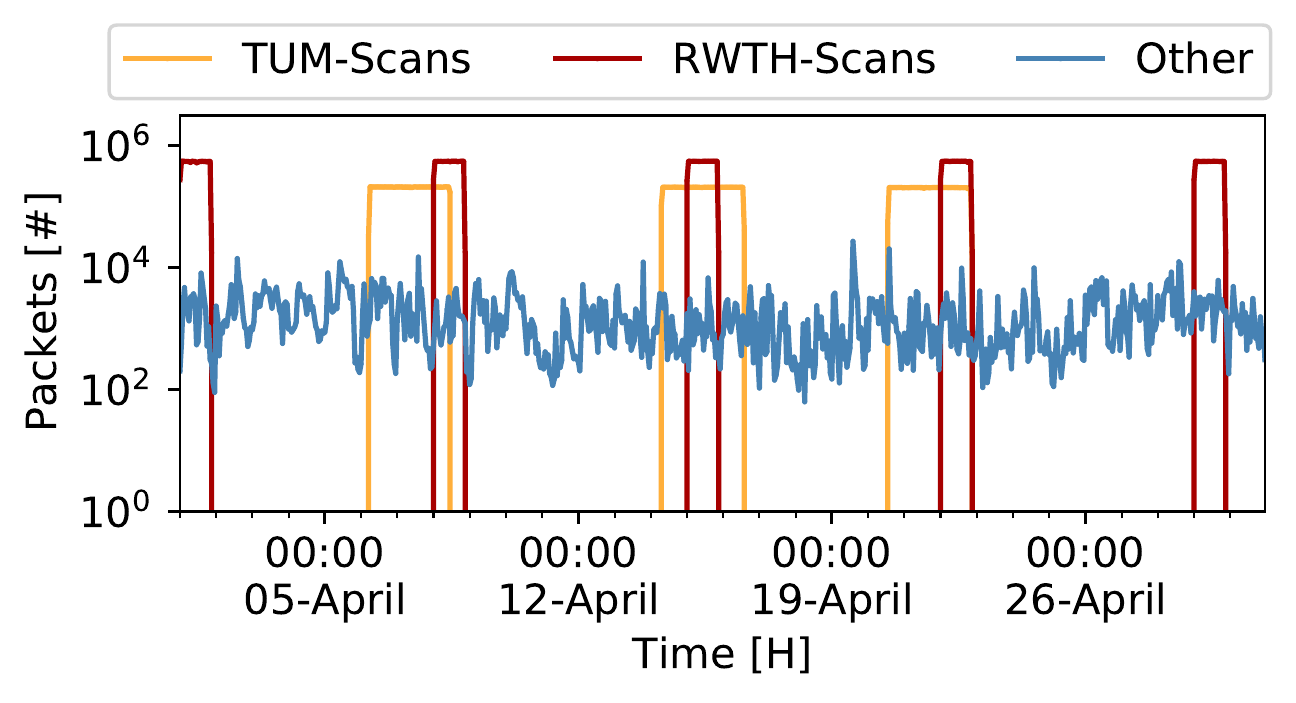}
      \caption{QUIC traffic seen at the UCSD network telescope. In the remaining analyses, we identify and remove the extreme bias of research scanners.}
  \label{fig:packets_with_scanners}
  \end{center}
\end{figure}

\paragraph{QUIC IBR is dominated by research scanners}
We first inspect the total packet count.
Since multiple QUIC messages can be embedded in a single IP/UDP packet, we emphasize that we count packets and not individual QUIC messages.
Overall, we observe 92 million QUIC packets in \mbox{April~1-30, 2021}.
This data set is dominated by periodic scans that target the complete IPv4 address space.
Each Internet-wide, single-packet scan sends $2^{23} \sim 8\times10^{6} $ packets to the telescope.
In total, 98.5\% of QUIC packets are generated by research projects from two universities, Technische Universität München (TUM) and RWTH Aachen Universität~(RWTH).
We show the amount of packets per hour and compare research scanners with other traffic sources in \autoref{fig:packets_with_scanners}.
Since these scanners clearly perform regular scans of the entire address space, we expect this bias also at other vantage points.
We remove traffic from research scanners in the subsequent analyses to focus on the \emph{other} traffic.

\begin{figure}[t]
  \begin{center}
  \includegraphics[width=1\columnwidth]{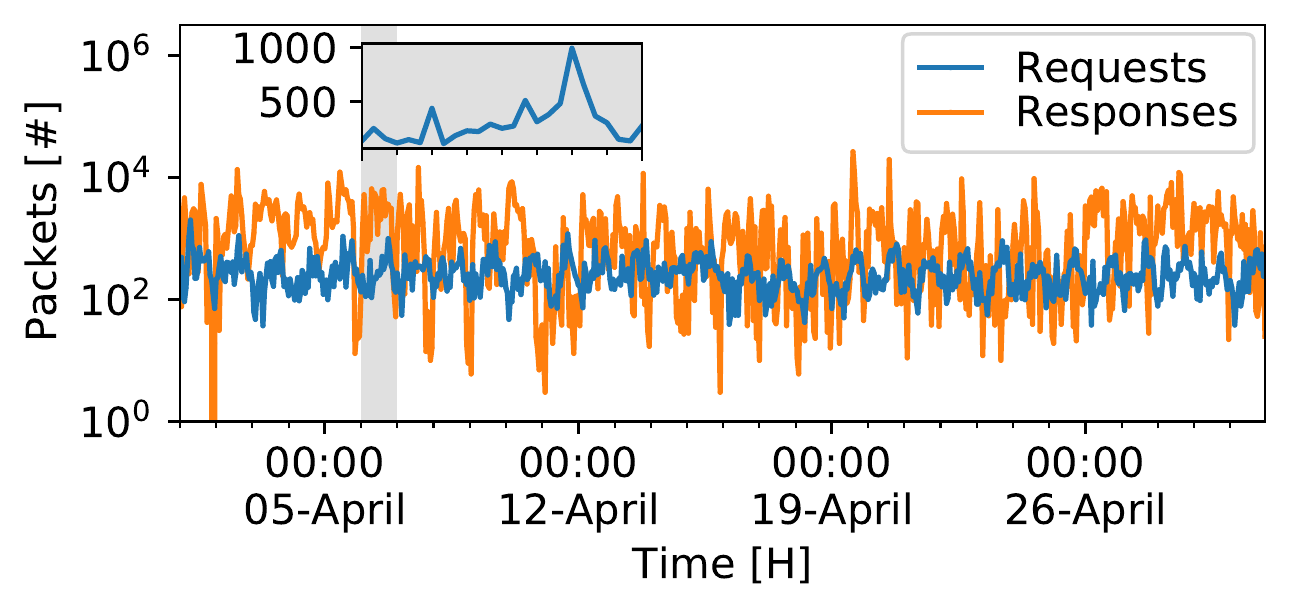}
      \caption{Number of QUIC packets by type.
      Requests exhibit rather stable, diurnal activities with peaks at 6:00am and 6:00pm~UTC (see insert for representative day).
      Responses are very erratic, hinting at flood events.}
  \label{fig:packets_req_resp}
  \end{center}
\end{figure}

\paragraph{QUIC requests follow diurnal patterns, responses are erratic}
In the sanitized traffic, we find 15\% QUIC requests and 85\% QUIC responses.
\autoref{fig:packets_req_resp} shows the number of requests and responses per hour.
Requests follow stable, diurnal patterns with peaks at 6:00am and 6:00pm UTC.
We demonstrate this with the inlet for a representative day, April 06, 2021.
Response traffic is very erratic, exhibiting high peaks and drops per event.
This behavior might hint at DoS events \cite{Balkanli2014Backscatter, Blenn2017Spectrum}, which we will inspect in more detail in \autoref{sec:analysis-dos-traffic}.

\paragraph{We can reuse established thresholds for QUIC sessions}
In order to move from a packet-based to an event-based perspective, we now group singular packets into sessions.
To this end, we aggregate packets using the source IP address and a \texttt{timeout} threshold,
\ie packets from a specific source belong to a single session as long as the inactivity period between them is no longer than the \texttt{timeout}.

\begin{figure}[b]
  \begin{center}
  \includegraphics[width=1\columnwidth]{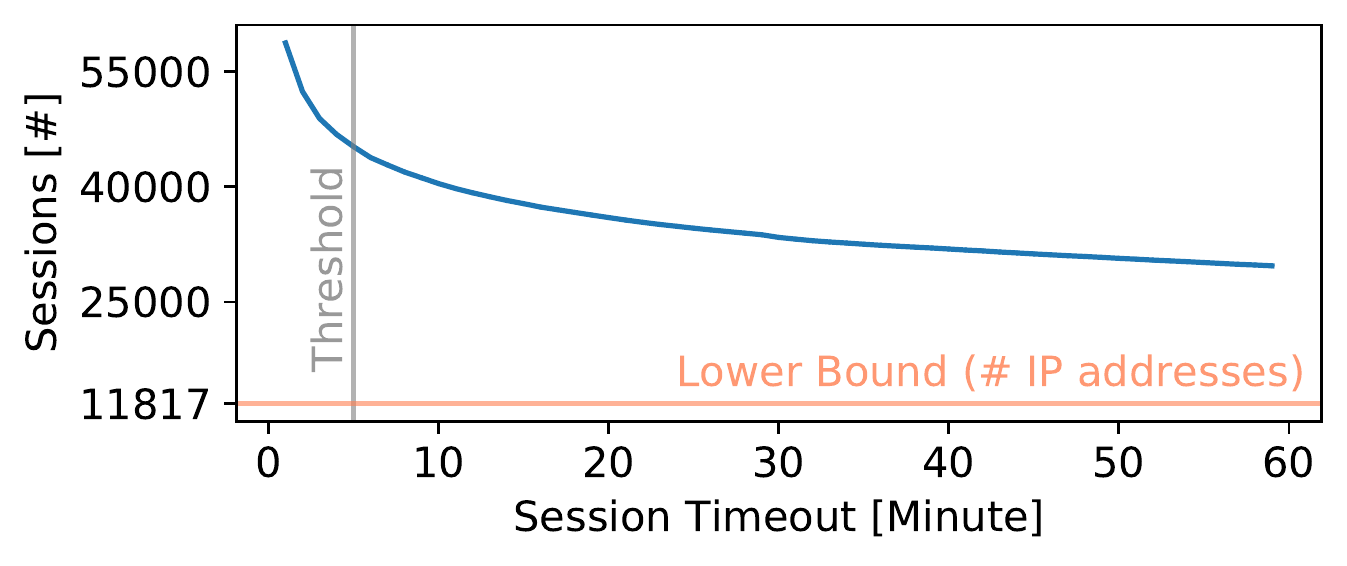}
      \caption{Influence of the timeout parameter on the number of sessions. We select 5~minutes as the final threshold to group correlated packets into sessions. }
  \label{fig:session_timeouts}
  \end{center}
\end{figure}

\autoref{fig:session_timeouts} exhibits the number of detected sessions given a timeout value between 1 and 60~minutes.
The lower bound of this plot is defined by \texttt{timeout=$\infty$}, which groups all packets of a source into a single event.
We see a significant reduction of sessions until $\sim$5~minutes, which is why we choose this knee point as our threshold.
This timeout is coherent with prior work that applied timeouts to IBR traffic~\citep{moore2006inferring, jonker2017millions}.

\subsection{QUIC DoS Traffic}
\label{sec:analysis-dos-traffic}

\paragraph{Request sessions are non-benign and originate from eyeball networks. Response sessions are DoS backscatter from content providers}
We now inspect and contextualize the observed sessions.
Overall, we find 18k sessions containing only requests, 26k sessions containing only responses.
We do not observe sessions with both packet types.
On average, a request session consists of 11~packets and response sessions of 44~packets.

\begin{figure}[t]
  \begin{center}
  \includegraphics[width=1\columnwidth]{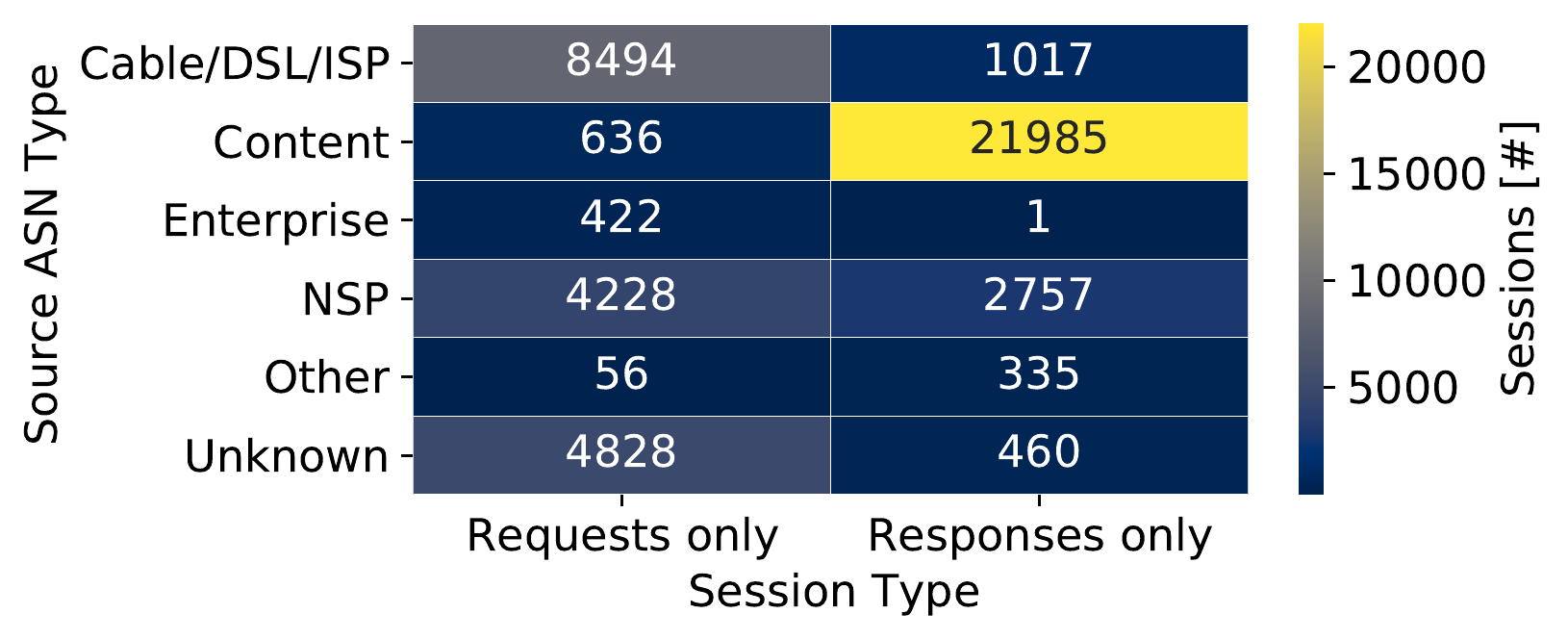}
      \caption{Source network types of sessions. Requests originate predominantly from eyeballs. Responses are received almost exclusively from content networks.}
  \label{fig:asn_types}
  \end{center}
\end{figure}

For each session, we map the source autonomous system number (ASN) and the AS network type using PeeringDB.
We find that request sessions originate from eyeball networks and that response sessions are received from content provider networks, see \autoref{fig:asn_types}.
This fits into the assumption that we \one receive malicious scans from bots hosted in eyeball networks (\eg Mirai \cite{avast-mirai-ongoing, aabbb-umb-17}) and \two~receive DoS backscatter from legitimate QUIC servers.

\begin{figure}[b]
  \begin{center}
  \includegraphics[width=1\columnwidth]{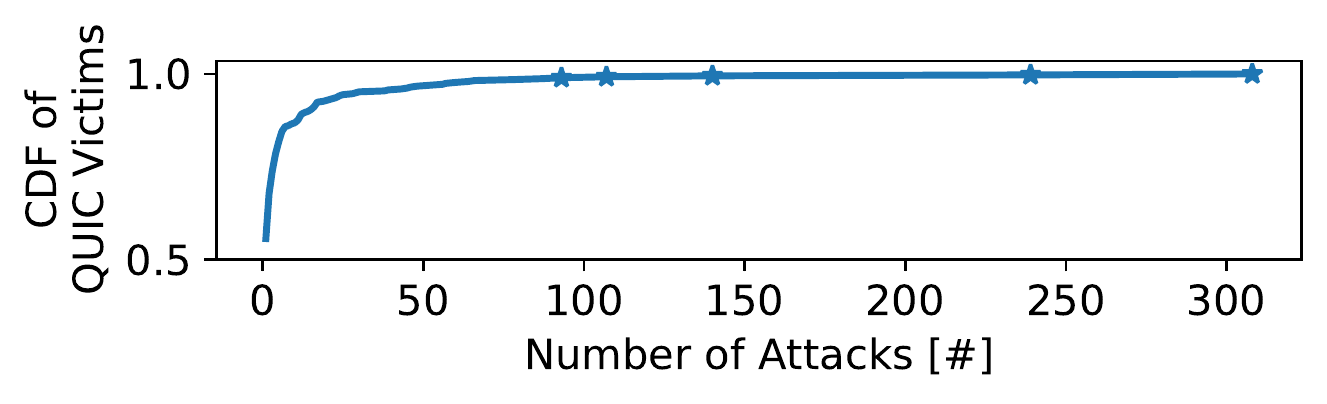}
      \caption{CDF for number of attacks per QUIC flood victim. More than half of the victims are only attacked once during our measurement period. Last 5 data points are highlighted.}
  \label{fig:cdf_reoccurence}
  \end{center}
\end{figure}

\begin{figure*}[t]
  \begin{center}
  \subfigure[Flood Durations\label{fig:cdf_duration}]{\includegraphics[width=1\columnwidth]{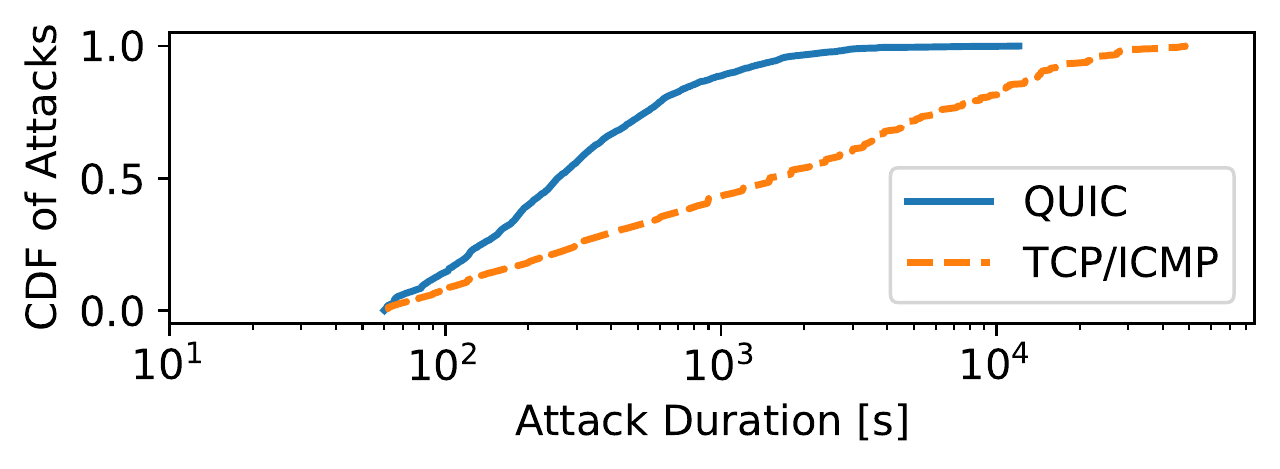}}
  \subfigure[Flood Intensities\label{fig:cdf_intensity}]{\includegraphics[width=1\columnwidth]{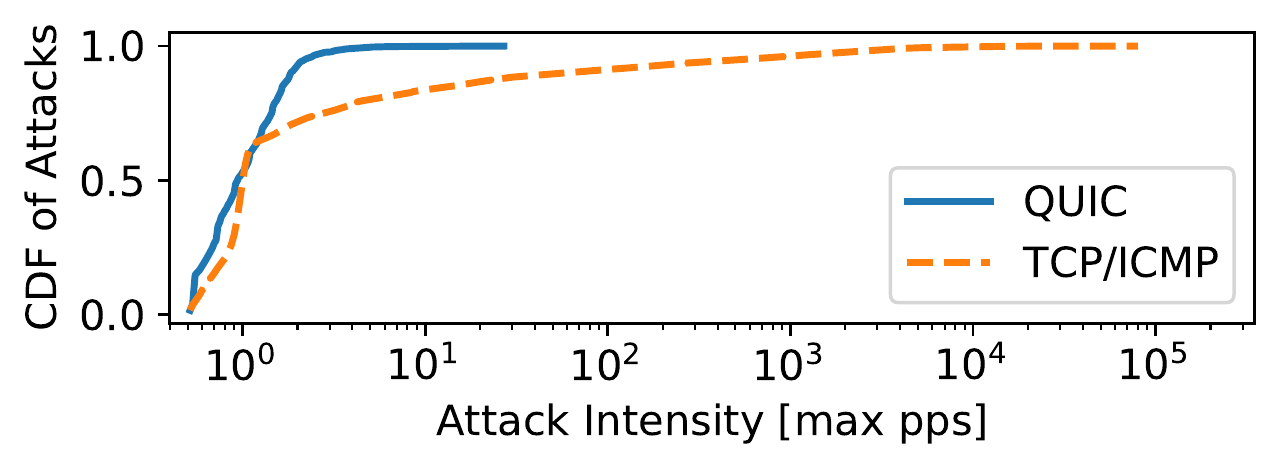}}
      \caption{CDF of flood durations and intensities, comparing QUIC and TCP/ICMP. QUIC floods are shorter but the median intensity of QUIC floods is as severe as for common backscatter events.}
  \label{fig:cdf_duration_intensity}
  \end{center}
\end{figure*}

Taking a closer look at response sessions we find traffic that exhibits common DoS patterns, using, again, session timeouts and thresholds defined by Moore \etal \citep{moore2006inferring}.
To identify attacks, we select backscatter sessions with \one more than 25 packets, \two a duration longer than 60 seconds, and \three a maximum packet rate of higher than 0.5pps, which is calculated over all 1-minutes slots of the respective event.
Finally, we find 2905 attacks which correspond to 11\% of all response sessions.
Attacks target a total of 394~unique victims, with more than half beeing attacked only once, compare \autoref{fig:cdf_reoccurence}.
By correlating the victims with data obtained from active scans~\cite{rueth2018quic}, we find that 98\% of attacks target well-known QUIC~servers.
We take a closer look at the low-volume backscatter sessions and our threshold configuration in \autoref{sec:dropped-backscatter}.

To bolster our observations we correlate the sources of request sessions with data from an reactive vantage point, the GreyNoise honeypot platform.
Using the GreyNoise classification, we do not find any signs of benign scanners and 2.3\% of the sources are tagged as known bruteforcers or part of a botnet such as Mirai or Eternalblue.
Most request sessions originate from Bangladesh~(34\%), USA~(27\%), and Algeria~(8\%).

\paragraph{QUIC floods are shorter but on average as severe as common TCP/ICMP floods}
We now compare QUIC DoS floods with floods for common protocols, \ie TCP and ICMP.
The duration and intensity of attacks has been shown by Jonker \etal \citep{jonker2017millions}.
To allow for comparison, we reproduce the analysis based on our current setup.
Overall, we find 282k attacks for common protocols.

\autoref{fig:cdf_duration} shows the distribution of attack durations.
QUIC floods are shorter than TCP/ICMP DoS attacks.
The median QUIC flood lasts 255~seconds, for TCP/ICMP protocols we observe 1499~seconds.
The median attack intensity, however, is close to 1 maximum packet per second (max pps) in both cases, see \autoref{fig:cdf_intensity}.
To estimate the traffic rate from the global Internet towards a victim, we may assume 512 $\times$ max pps since the UCSD~telescope covers 1/512 of the total IPv4 address space.

\begin{figure}[b]
  \begin{center}
  \includegraphics[width=1\columnwidth]{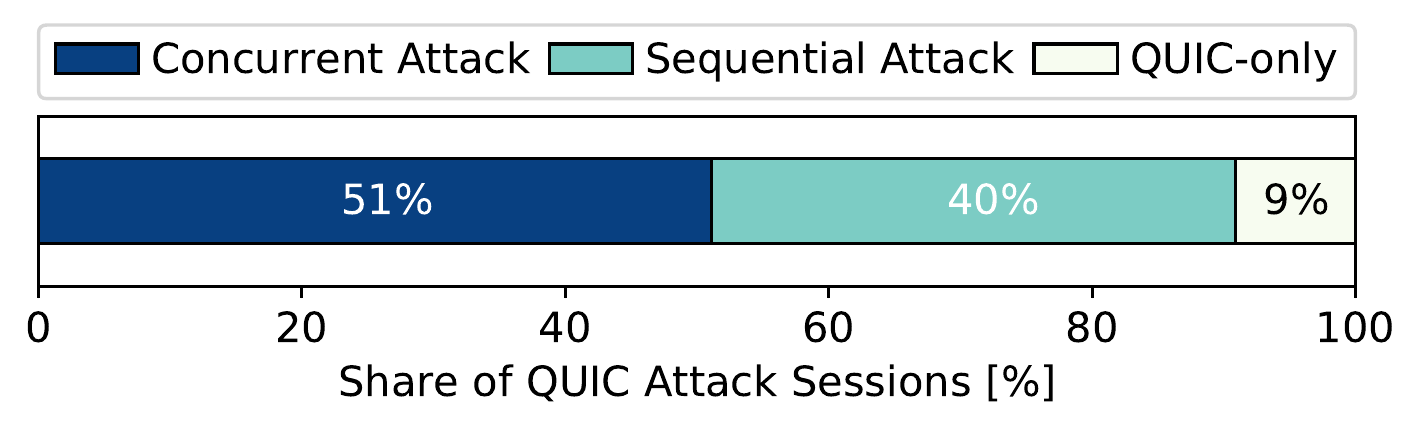}
      \caption{Multi-vector attacks: Half of the QUIC attacks occur concurrently with TCP/ICMP floods.}
  \label{fig:attack_overlap}
  \end{center}
\end{figure}

\paragraph{QUIC floods are part of multi-vector attacks and highly correlated with TCP/ICMP floods}
So far, we looked at QUIC floods in isolation.
We now check whether they are part of a larger multi-vector attack towards a single victim. %
To our surprise, 51\% of QUIC floods overlap in time with common (TCP/ICMP) DoS floods (see \emph{concurrent} attacks in \autoref{fig:attack_overlap}).
We require attacks to overlap for at least one~second to label them as \emph{concurrent}.
Another 40\% of QUIC floods target a victim in \emph{sequence}, \ie the victim was also attacked by a TCP or ICMP flood during our measurement period but at a different time.
In such cases, the gap between a QUIC and the nearest TCP/ICMP attack is 36 hours on average.
More details about concurrent and sequential attacks, including an example, are presented in \autoref{sec:multi-vector}.
Only 9\% of QUIC attacks do not relate to any TCP/ICMP event.

We argue that the reasons for multi-vector attacks are twofold.
First, multi-vector attacks are harder to detect as they keep the traffic volumes for each attack vector low.
Second, multi-vector attacks are more difficult to mitigate since they require more complex filter rules.

\begin{figure*}[t]
  \begin{center}
  \includegraphics[width=2.1\columnwidth]{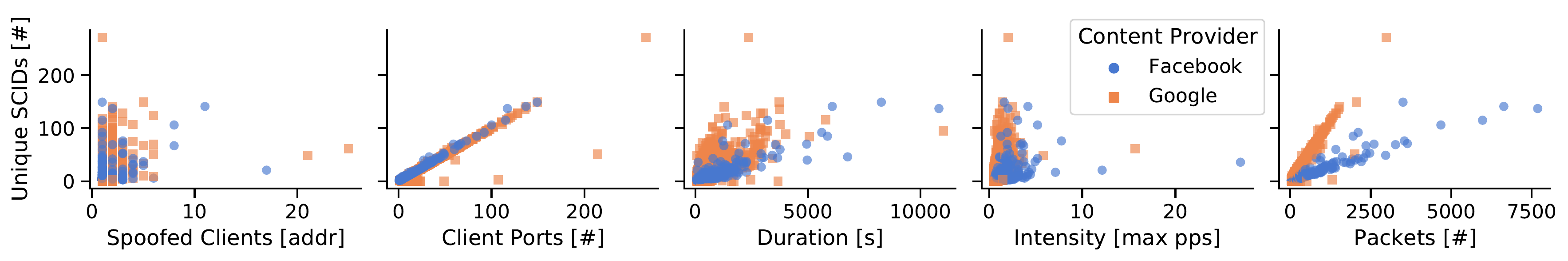}
      \caption{>83\% of attacks target two content providers.  QUIC floods utilize multiple client addresses and ports. %
      Despite a lower packet count, Google reacts with more SCIDs, indicating higher server load.}
  \label{fig:pairplot_attacks}
  \end{center}
\end{figure*}

\paragraph{Well-known content providers are under attack}
We find that 58\% of attacks target Google and 25\% target Facebook.
\autoref{fig:pairplot_attacks} compares attack properties for these two content providers
to showcase potential differences in attack patterns across content providers and their deployed QUIC variants.
We consider SCIDs in the backscatter traffic.
The SCID is a QUIC-specific feature, which may serve to assess allocated resources because a context is reserved for each unique connection.
We do not show the number of DCIDs in this figure, since they are not available in our backscatter traffic and they are not required to route to the correct endpoint \cite{RFC-9000}. 
We carefully checked that the packets are valid, though, by verifying that the DCID length attribute is set to zero.%

Overall, the number of spoofed client IP addresses is relatively low.
The randomization of client ports, however, is the driving factor for new SCIDs at the attacked server.
Despite a lower packet count, Google reacts with more SCIDs, which indicates a higher server load.
We observe QUIC variant \texttt{mvfst-draft-27} (95\%) for Facebook attacks, and \texttt{draft-29} (78\%) for Google. %

These results suggest that operators may protect against QUIC floods by filtering based on common transport protocol features (\ie ports) instead of using QUIC-specific features (\ie SCIDS), which eases the deployment of countermeasures.

\section{Discussion and Outlook}
\label{sec:discussion}

\paragraph{Attack patterns are valid}
Captured QUIC events that are suspect to DoS consist of 31\% \texttt{Initial} and 57\% \texttt{Handshake} messages on average.
The \texttt{Initial} messages we observe do not contain an (unencrypted) TLS \texttt{Client Hello} and thus can be attributed to (encrypted) \texttt{Server Hello} replies. 
Hence, these packets match the backscatter pattern of a QUIC attack (see Section~\ref{sec:related_work}). QUIC sends multiple UDP packets in response to the \texttt{Initial} packet:
The first packet contains one \texttt{Initial} QUIC packet carrying the \texttt{Server Hello} and one encrypted \texttt{Handshake} message followed by a second datagram with a single \texttt{Handshake} message.
The ratio in these packets roughly matches the ratio of one third \texttt{Initial} packets and two thirds \texttt{Handshake} messages.

\paragraph{\texttt{RETRY} attack mitigation is not deployed}
We did not capture any \texttt{RETRY} messages.
The absence of \texttt{RETRY} indicates the lack of deployment of a defense mechanism.
To validate this observation, we select the ten most frequently attacked servers from Google and Facebook.
When actively connecting to these servers with a QUIC client we also do not receive any \texttt{RETRY} messages.
This supports the observations that we made based on data in the telescope.

Although the QUIC implementations of Google and Facebook support \texttt{RETRY} messages (Google since mid 2019 \cite{commitGoogleRetry}, Facebooks since the end of 2020 \cite{commitFacebookRetry}), it looks like these content providers deliberately decide to not utilize this feature.
This decision is potentially due to the performance penalty of \texttt{RETRY} messages.
However, for frequently utilized services as in the case of large content providers, this penalty could be alleviated by the session resumption feature in QUIC.
Also, \texttt{RETRYs} could be deployed adaptively and only used when high load occurs.

\paragraph{QUIC servers quickly experience DoS}
To check whether the observed QUIC floods can be attributed to DoS attacks we experimentally analyzed the impact of QUIC handshakes on a common web server implementation.
For this, we set up NGINX on a modern 128-core machine with 512 GB of RAM, which connects to a client via Gigabit Ethernet.
NGINX supports the QUIC RFC \cite{RFC-9000} and eBPF optimizations~\cite{nginx2021ebpf}.
We record ~{500,000} packets using the QUIC~client \texttt{quiche}, version 0.9.0, Cloudflare's QUIC reference implementation.
To simulate attacks, we then replay \emph{only} client \texttt{Initial} messages at varying packet rates towards new server instances.
Replaying avoids bias from hand-crafting QUIC packets. 

Our results are summarized in Table~\ref{tab:dos_nginx} alongside the configuration (we use 1024 client connections per worker which is twice the default).
Since each request elicits four datagrams in response (two datagrams with \texttt{Initial} and \texttt{Handshake} packets plus two keep-alive \texttt{PING} packets after a short delay) we expect four times as many server responses.
To determine how many requests were answered we match the respective DCIDs and SCIDs and calculate the service availability ratio.

With four worker processes even small packet rates can lead to significantly reduced service availability.
At 1,000 probes per second (pps) only 7\% of our requests are answered.
In auto-mode, NGINX deploys 128 workers, but larger attack volumes of 10,000~pps continue to impact its availability.
Note that these attacks do not impair the general availability of the machine but focus on the service.
Extrapolating our observed 27 pps from a \texttt{/9} IP prefix to the size of the Internet leads us to believe that attacks of more than {10,000} pps (27 $\cdot$ 512=13,824) are ongoing.
In our benchmarks, \texttt{RETRY} packets successfully mitigate these attacks but they come at the high cost of an additional round trip.

The vulnerability based on the QUIC stateful handshake is not specific to an im\-ple\-men\-ta\-tion, but relates to the protocol design. %
We expect all implementations to be prone to this attack type, the exact pps rates might vary, though.
A recent benchmark of picoquic~\cite{Huitema2020DDoS} also observed DoS at around {10,000} pps but a successful attack mitigation with \texttt{RETRYs}.
Latest interoperability tests \cite{quic2021interop} show that the majority of QUIC server and client implementations correctly support the \texttt{RETRY}~option. 
This leaves the decision on robustness versus speed up to the service providers.

\setlength{\tabcolsep}{1.60pt}
\renewcommand{\arraystretch}{0.8}
\begin{table}
  \caption{Tests on a local NGINX instance show that the backscatter volume we observed can significantly impact the responsiveness of the web server.}
  \label{tab:dos_nginx}
  \centering
  \begin{tabular}{>{\raggedleft\arraybackslash}p{1.25cm} >{\centering\arraybackslash}p{0.90cm} >{\centering\arraybackslash}p{1.30cm} >{\raggedleft\arraybackslash}p{1.15cm} >{\raggedleft\arraybackslash}p{1.25cm} >{\raggedleft\arraybackslash}p{1.4cm} >{\centering\arraybackslash}p{0.8cm} }
    \toprule
    \centering{Attack} & \multicolumn{2}{c}{NGINX Config} & \multicolumn{4}{c}{Results} \\
    \cmidrule(r){1-1}
    \cmidrule{2-3}
    \cmidrule(l){4-7}

    Volume [pps] & QUIC Retry  & Workers [\#] & Client [\# Req] & Server [\# Resp] & Service Available & Extra RTT %
    \\

    \midrule
    \rowcolor{ColorAtkNo!40}
        {10} & \xmark &    4 &   {3,001} &   {12,004} & 100\% & \xmark %
    \\
    \rowcolor{ColorAtkYes!40}
       {100} & \xmark &    4 &  {30,001} &   {81,680} &  68\% & \xmark %
    \\
    \rowcolor{ColorAtkYes!40}
      {1,000} & \xmark &    4 & {300,001} &   {81,680} &   7\% & \xmark %
    \\
    \midrule
    \rowcolor{ColorAtkNo!40}
      {1,000} & \xmark &    auto=128 & {300,001} & {1,200,004} & 100\% & \xmark %
    \\
    \rowcolor{ColorAtkYes!40}
     {10,000} & \xmark &    auto=128 & {500,000} &  {522,752} &  26\% & \xmark %
    \\ 
    \rowcolor{ColorAtkYes!40}
    {100,000} & \xmark &    auto=128 & {498,991} &  {322,158} &  26\% & \xmark %
    \\
    \midrule
    \rowcolor{ColorAtkNo!40}
      {1,000} & \cmark & 4 & {300,001} &  {300,001} & 100\% & \cmark %
    \\
    \rowcolor{ColorAtkNo!40}
     {10,000} & \cmark & 4 & {500,000} &  {500,000} & 100\% & \cmark %
    \\ 
    \rowcolor{ColorAtkNo!40}
     {100,000} & \cmark & 4 & {500,000} &  {500,000} & 100\% & \cmark %
    \\ 
    
    \bottomrule
  \end{tabular}
\end{table}

\paragraph{Attack duration and intensity}
We found that QUIC floods are shorter but the median max packet rate is similar compared to TCP and ICMP.
The max packet rate is an indicator of the attack intensity but it also reflects the capability of a victim to sustain under load---for well-provisioned victims we likely observe higher rates as those victims are still able to send data.
This also applies to the observed durations. %
Backscatter events can stop for various reasons: \one the attack has ended, \two a mitigation was initiated, or \three the attacked service is completely unresponsive.
Hence, shorter durations might indicate that QUIC attacks lead to a faster resource exhaustion compared to common protocols.
Analyzing this, \eg by using reactive scans or correlating with other data, will be part of our future work.

\begin{acks}
We would like to thank our shepherd Jelena Mirkovic and the anonymous reviewers for their helpful feedback.
This work was partly supported by the \grantsponsor{BMBF}{German Federal Ministry of Education and Research (BMBF)}{https://www.bmbf.de/} within the project \grantnum{BMBF}{PRIMEnet}.
\end{acks}

\label{lastpage}

\bibliographystyle{ACM-Reference-Format}
\bibliography{bibliography}

\begin{appendix}
  \balance
\section{Artifacts}
\label{sec:artifacts}

We support reproducible research.
All artifacts of this paper are available on \url{https://doi.org/10.5281/zenodo.5504168}.

\section{Non-Attack Backscatter and Threshold Configuration}
\label{sec:dropped-backscatter}

We reused thresholds defined by Moore \etal \cite{moore2006inferring} to infer DoS events.
We classified 11\% of response sessions as attacks.
Although this seems like an extreme reduction of events, we argue that underestimation is better than overestimation, because this prevents false positives.
Nevertheless, we also checked the excluded response sessions for deviating trends.
We do not find any anomalies.
Excluded events have a median intensity of 0.18~max~pps, a duration of 7~seconds, and consist of 11 packets.
Such low-volume events point to misconfigurations and are most likely insignificant for our DoS analysis, hence, should be excluded.

To further understand the effects of our threshold configuration, we introduce the threshold weight $w$.
We multiply each threshold by $w$, which leads to a more relaxed ($w<1$) or stricter ($w>1$) attack detection.
If  $w=1$, the default threshold configuration as defined by Moore \etal~\cite{moore2006inferring} is used.

\begin{figure}[h]
  \begin{center}
  \includegraphics[width=1\columnwidth]{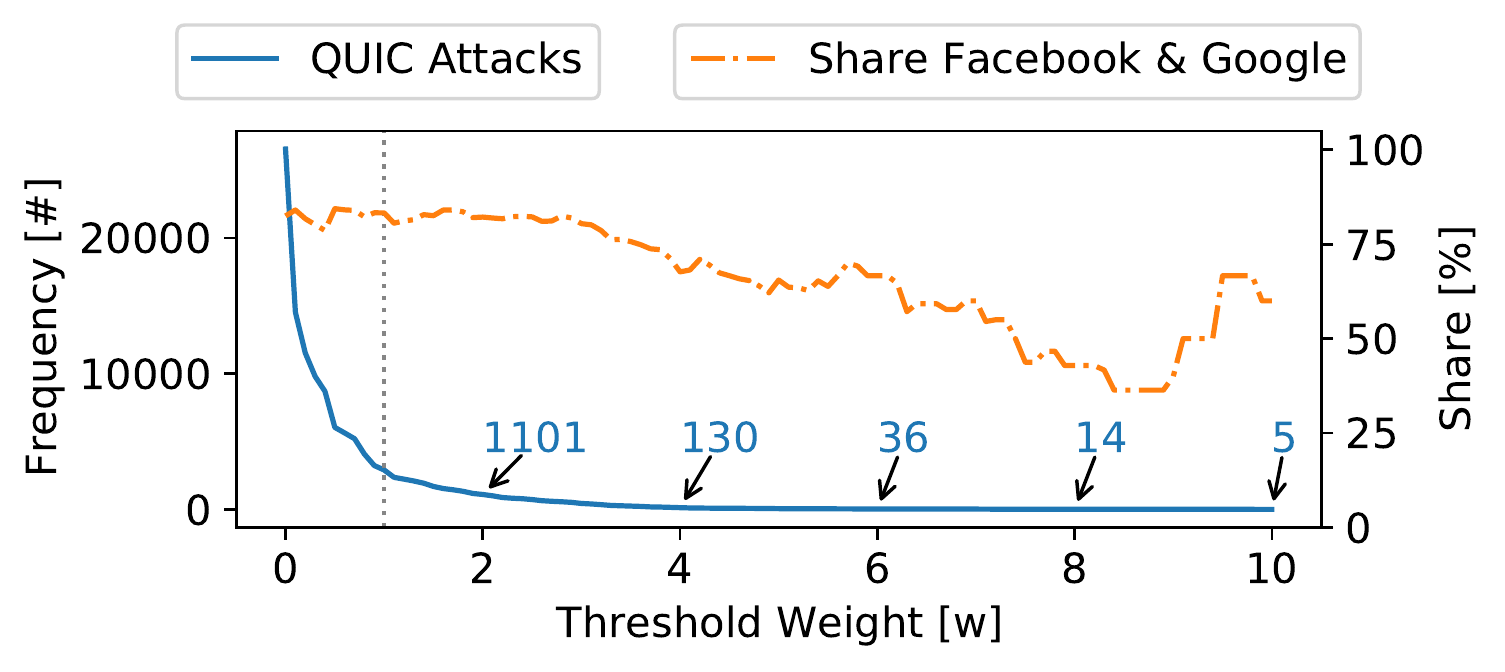}
      \caption{Varying the DoS threshold defined by Moore \etal~\cite{moore2006inferring} to show the impact on the number of detected attacks and the relative share of affected content infrastructures. Even for a very strict threshold configuration of $w=10$, we find QUIC attacks.}
  \label{fig:ts_inc_thresholds_merged}
  \end{center}
\end{figure}

\autoref{fig:ts_inc_thresholds_merged} shows the number of detected attacks.
We exclude many low-volume backscatter events for $w\le0.3$, but even for an extreme configuration of $w=10$ we still classify five backscatter sessions as attacks.
On the secondary $y$-axis, we show that the share of well-known content providers remains high independently of $w$.
These results bolster our main insight that QUIC \texttt{Initial} floods are used to attack large content providers.

\section{Details about Attacks}
\label{sec:multi-vector}
We introduced concurrent (\ie multi-vector) and sequential attacks in \autoref{sec:analysis-dos-traffic}.
In this section, we illustrate those attacks based on a concrete example and present more details about the time overlap between concurrent QUIC and TCP/ICMP attacks, as well as the time gap between sequential attacks.

\subsection{Illustration of Multi-vector versus Sequential Attacks}

\begin{figure}[h]
  \begin{center}
  \includegraphics[width=1\columnwidth]{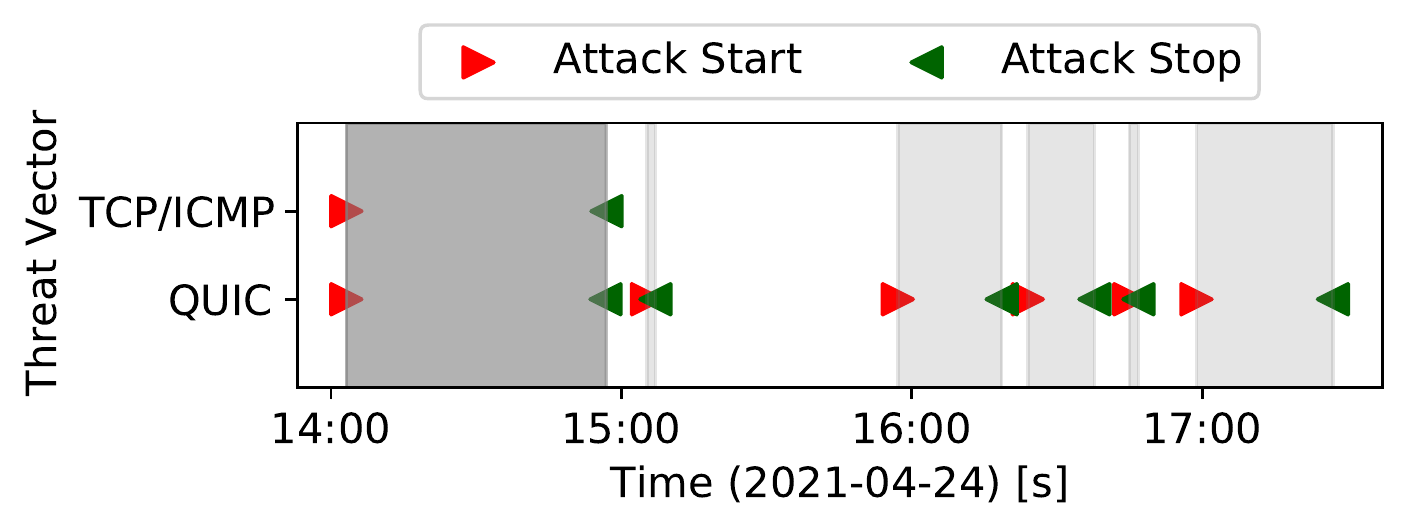}
      \caption{Attacks towards a single victim. We observe one concurrent usage of attack vectors, \ie a multi-vector attack, followed by five sequential QUIC floods.}
  \label{fig:fb_attack_concurrency}
  \end{center}
\end{figure}

In \autoref{fig:fb_attack_concurrency}, we illustrate concurrent and sequential attacks based on a snapshot for a single victim.
First, the victim is attacked by one QUIC and one TCP/ICMP attack that take place at the same time (or \emph{concurrently}).
Please note that these two attacks have an almost perfect overlap but we classify any two attacks as concurrent if the respective time ranges overlap in at least a single time unit, \ie they share at least one mutual second.
Such a perfect overlap is very likely, compare \autoref{sec:attack-overlap}.
We then observe five QUIC attacks in \emph{sequence} to the first TCP/ICMP attack.

\subsection{Overlap of Concurrent Attacks}
\label{sec:attack-overlap}

\begin{figure}[h]
  \begin{center}
  \includegraphics[width=1\columnwidth]{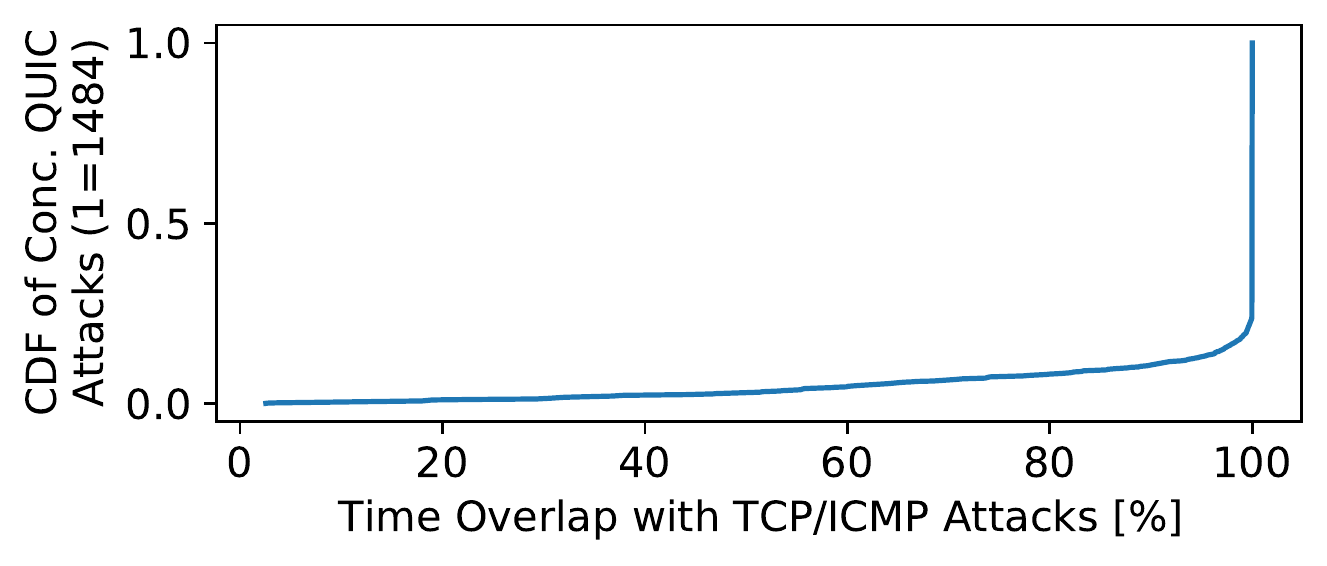}
      \caption{Attack overlap of multi-vector attacks. Most concurrent QUIC attacks almost completely overlap with attacks that use common protocols.}
  \label{fig:cdf_interval_overlap}
  \end{center}
\end{figure}

We now investigate how QUIC attacks overlap with common TCP or ICMP attacks.
To this end, we calculate the share of overlapping seconds for each QUIC attack that is part of a concurrent attack.
\autoref{fig:cdf_interval_overlap} shows the distribution.
We find a high correlation between QUIC and TCP/ICMP attacks.
Three quarters of all concurrent QUIC attacks occur completely in parallel to an TCP/ICMP attack (100\% in the CDF).
On average, multi-vector QUIC~attacks share 95\% of the attack time with common attacks.

\clearpage

\subsection{Time Gaps Between Sequential QUIC Attacks and TCP/ICMP Attacks}
\begin{figure}[h]
  \begin{center}
  \includegraphics[width=1\columnwidth]{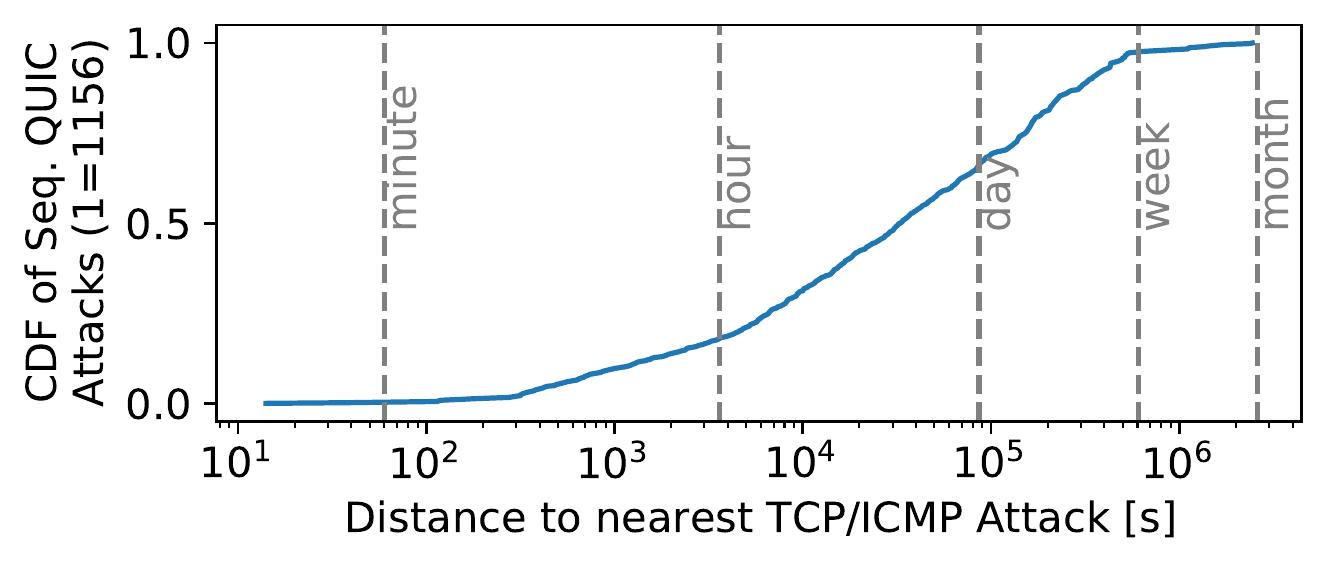}
      \caption{Distribution of time gaps between the end (or start) of a sequential QUIC attack and the start (or end) of a TCP/ICMP attack.}
  \label{fig:cdf_dist_seq}
  \end{center}
\end{figure}

We label an attack session \emph{sequential attack} when we observe QUIC and TCP/ICMP attack traffic to same victim but QUIC and TCP/ICMP attacks do not overlap.
\autoref{fig:cdf_dist_seq} exhibits the distribution of time gaps between both attack vectors.
There is a break of more than one~hour for 82\% of the sequential attacks.
In some cases, a break may take up to 28~days.
These long time gaps suggest that sequential attacks are indeed not part of a multi-vector attack.

\end{appendix}

\end{document}